\documentclass[journal,10pt]{IEEEtran}

\usepackage{amsfonts}
\usepackage{amssymb}
\usepackage{mathtools,amsmath}
\usepackage{amsthm}
\usepackage{bm}
\usepackage{graphicx}
\usepackage{multirow}
\usepackage{color}
\usepackage{graphics}
\usepackage{cite}
\usepackage{enumitem}
\usepackage{caption}
\usepackage{listings}
\usepackage[nottoc,numbib]{tocbibind}
\usepackage{mathtools}
\usepackage{rotating}
\usepackage{algorithm}
\usepackage{algpseudocode}
\usepackage{nccmath}

\newcommand{\beq}{\begin{equation}}
\newcommand{\eeq}{\end{equation}}
\def\be#1\ee{\begin{align}#1\end{align}}
\newcommand{\ov } {\over }

\label{CryptosystemDefinition}

\newtheorem*{comp}{Comparison Operation}
\newtheorem*{select}{Selection Operations}

\begin{document}

\title{Conditionals in Homomorphic Encryption, and Machine Learning Applications}

\author{Diego Chialva
        and~Ann Dooms
\thanks{D. Chialva is with the ERCEA (European Research Council Executive Agency) and A. Dooms is with the Department of Mathematics, Vrije Universiteit Brussel, Pleinlaan 2, 1050 Brussels, Belgium.

\textbf{Disclaimer.} The views expressed in this paper by Diego Chialva are the author's. They do not necessarily reflect the views or official positions of the European Commission, the European Research Council Executive Agency 
or the ERC Scientific Council.}

}

\maketitle

\begin{abstract}

Homomorphic encryption aims at allowing computations on encrypted data without decryption other than that of the final result. This could provide an elegant solution to the issue of privacy preservation in data-based applications, such as those using machine learning, but several open issues hamper this plan. In this work we assess the possibility for homomorphic encryption to fully implement its program without relying on other techniques, such as multiparty computation (SMPC), which may be impossible in many use cases (for instance due to the high level of communication required). We proceed in two steps: i) on the basis of the structured program theorem [Bohm, Jacopini] we identify the relevant minimal set of operations homomorphic encryption must be able to perform to implement any algorithm; and ii) we analyse the possibility to solve -and propose an implementation for- the most fundamentally relevant issue as it emerges from our analysis, that is, the implementation of conditionals (requiring comparison and selection/jump operations). We show how this issue clashes with the fundamental requirements of homomorphic encryption and could represent a drawback for its use as a complete solution for privacy preservation in data-based applications, in particular machine learning.  Our approach for comparisons is novel and entirely embedded in homomorphic encryption, while previous studies relied on other techniques, such as SMPC, demanding high level of communication among parties, and decryption of intermediate results from data-owners. A number of studies have indeed dealt with comparisons, but typically their algorithms rely on other techniques, such as secure multiparty computation, which required a) high level of communication among parties, and b) the data owner to decrypt intermediate results. Our protocol is also provably safe (sharing the same safety as the homomorphic encryption schemes), differently from other techniques such as Order-Preserving/Revealing-Encryption (OPE/ORE).
\end{abstract}


\section{Introduction}\label{IntroductionSect}

Machine learning, data mining and predictive data analytics
represent an ensemble of techniques and algorithms (which for simplicity we will in the following indicate simply as "machine learning") that allow systems to act and make predictions without being explicitly programmed in full detail to do so, but by leveraging their input data with inference techniques. They have nowadays an overwhelming number of practical applications providing us with an unprecedented level of comfort and services, 
from tailored suggestion systems, to ``personalised medicine'', and several other services. 

However, these advantages typically come at the price of loosing individual privacy, as personal or valuable information is used by the algorithms and the third parties operating them. 
This issue has spawn research activity at different levels.  
Very roughly speaking we can divide the developed privacy-preservation techniques in two classes: those that work by modifying the data themselves and those that modify the representation of the data, but not the actual data content.

Techniques of the first class act on the datasets holding the privacy-concerned data and can be divided in a few subclasses \cite{AY2008}. Common to all of them is the distinction between identifier, quasi-identifier and anonymous data.
Such techniques require only comparatively minor (conceptual) changes to the application algorithms acting on the data, but they have significant drawbacks. Indeed, they impose a trade-off between the degree of preserved privacy and the usefulness of the data: a ``privacy budget'', which has been shown to be quite limited \cite{FLJLPR2014}.  Moreover, such techniques appear to be beatable by the algorithms themselves and  database crossing attacks (that is, the use by an attacker of other, public or stolen, databases to ``complete'' or infer the relevant distorted information in the database of interest) \cite{FLJLPR2014}.

The other class of techniques has been proposed within cryptography.
Among the different research lines we recall:
secure multi-party computation, functional encryption, program obfuscation and homomorphic encryption, see for instance \cite{W2014, BSW2011, BGIRSVY2001, GGHRSW2013}.
These approaches differ in several aspects, including the set of functions that can be computed on the encrypted data and stage of development.

Homomorphic encryption aims at enabling the computation of arbitrary (in the case of \emph{fully homomorphic encryption}) or classes (in the case of \emph{partial homomorphic encryption}) of functions on encrypted data without having the need to decrypt them first and {\em limiting decryption to the very final result only}. This is in particular interesting for privacy preservation (including algorithm protection) in learning applications, and has been actively pursued in the latest years, e.g \cite{GLN2012, BBBCIV2017,  BLN2014, GDLLNW2016, AEH2015}. 
However, several open issues make homomorphic cryptosystems still unsuited for the vast majority of machine learning algorithms. 
Those that have been identified in the literature mainly are:
memory footprint, computational complexity, limited representable data (only integers and finite precision floats) and a restricted set of operations (only polynomial operations, that is addition and multiplication).

Such problems can however be divided into two classes.
The first class comprises issues such as the memory footprint and the computational complexity, which could be hoped to be trivially solved by technological (hardware) advancements, similarly to what has happened in deep learning. 
On the other hand, the second class of the above-listed problems, like the limited types of representable data and the lack of more general operations, must find a solution at the theoretical and cryptography level. It is this second class of issues that we are interested in within this work.

We therefore individuate and analyse the minimum set of basic operations necessary to implement any algorithm, whatever its complexity, and assess the possibility/impossibility to implement them in homomorphic encryption from first principles. 

We proceed on the basis of the well-known \emph{structured program theorem} \cite{BohmJacopini}, which states that every computable function\footnote{Technically, representable as a flow chart, such as all machine learning algorithms.} can be implemented in a programming language that combines subprograms in only three specific ways:
\begin{enumerate}
\item Executing one subprogram, and then another subprogram (sequence);
\item Executing one of two subprograms according to the value of a Boolean variable (selection);
\item Executing a subprogram until a Boolean variable is true (iteration).
\end{enumerate}

We observe that 2) and 3) require being able to perform \textit{conditionals}, that is \textit{comparison} operations to compare values and evaluate down to a single Boolean value, \textit{and selection/jump} operations to pick up the correct branch of a program. Hence in order to assess the possibility for homomorphic encryption to accommodate all algorithms (in particular machine learning ones), comparisons and selections/jumps must be implementable\footnote{Proving the impossibility of implementing such operations, would entail the impossibility to implement complex algorithms under homomorphic encryption. Clearly, in the opposite case, where the fundamental analysis is positive, one still needs to assess the effectiveness of the implementation, which may still condemn homomorphic encryption to be impractical.}.

In this work we address both the issue of comparison and of selection/jump operations.
Concerning the former, several proposals have been made concerning comparison operations in an encrypted setting, but not yet totally within
an homomorphic cryptosystem. There is even a claim that comparison would not be feasible in pure homomorphic encryption, see for example the comments in \cite{ABCGJRS2015}. 
Typical, well-studied approaches have been 
\begin{enumerate}
\itemsep=-1pt
 \item the so-called \textit{Order-Preserving-Encryption} (OPE) and its variants such as \textit{Modular-OPE} and \textit{Order-Revealing-Encryption} (ORE), see for instance \cite{AKSX2004, BCLN2009, BCN2011, PLZ2013, MCAKC2015, LW2016}, which do not belong to the homomorphic encryption class and, more seriously, have been proven to be not secure (for recent proofs, see for instance \cite{BDC2016, GSBNR2017}); 
\item \textit{secure multi-party computation} (SMPC) (for recent works see \cite{LJLA2017, DSZ2015}), which, although secure, require a high level of communication between parties (each single comparison in a machine learning training and prediction process must be performed by exchanging several messages), which may not be always possible; 
 \item combinations of homomorphic encryption with other cryptographic techniques such as SMPC to perform the comparisons \cite{EFGKLT2009, BATG2015, MZ2017, JVC2018, TPICOMM2014}. Again, these approaches do not manage to perform the comparisons exclusively on encrypted messages, as the data owner is required by the protocol to decrypt intermediate results, extract the significant bits for the comparison, re-encrypt and send the result back to the other party for the accomplishment of the algorithm. Such ``decryption in the middle'' hampers the purpose of homomorphic encryption (also, the need for a high level of communication between parties due to the use of SMPC may be impossible in a number of actual practical use cases).
\end{enumerate}

In this paper we develop, through a new approach, a technique to achieve comparisons in  homomorphic encryption (that is, with no need for communication between parties and acting exclusively on encrypted messages with no need for intermediate nor partial decryption).

When turning however to selection/jump operations, which are integral elements of conditionals, and thus of practically relevant algorithms, we will show that 
one hits a rather fundamental issue in homomorphic cryptosystems, namely the cryptographic requirement of \textit{semantic security}. 
We will show how a more limited form of selection/jump operations can still be implemented, and we discuss the limitation the above-mentioned issue imposes on the implementation of full machine learning algorithms.
In particular, this could represent a serious drawback in using homomorphic encryption for data analysis applications and for implementing algorithms in general, and could force to revisit (or abandon) that plan in its more ambitious formulation.

The article is organised as follows: we provide a brief introduction to homomorphic encryption in Section \ref{HomEncSect}, after which we study  comparisons and selections in a homomorphic setting in Section \ref{AnlResSect}. In Section \ref{EmpiricalResultsComparisonSect}
we present our methodology and test results. 
We finally discuss applications to machine learning in Sections \ref{ApplicationsSect}, \ref{MachineLearningSect}, 
where we highlight and provide precise and exhaustive examples of how the fundamental, general issues of homomorphic encryption that our analysis has revealed impact  the program of implementing machine learning algorithms in such framework. In our conclusions we also briefly touch upon the consequences of our work in applications different than machine learning.

\section{Homomorphic encryption}\label{HomEncSect}

A cryptosystem consists of three sets, a plaintext $P$, ciphertext $C$ and key space $K$, together with a family of encryption functions $\texttt{Encr}: K\times P\rightarrow C$ and decryption functions $\texttt{Decr}: K\times C\rightarrow P$ such that for each $k\in K$, there exists a $k'\in K$ such that $\texttt{Decr}( k', \texttt{Encr}( k, p )) = p$ for all $p\in P$.
Although in the literature $\texttt{Encr}$ is called encryption {\em function}, it is not exactly a function in the strict mathematical sense for most of the encryption schemes, because an element of (pseudo)randomness is involved such that applying it more than one time to the same key and plaintext, one obtains different ciphertexts.
Such probabilistic encryption schemes are favoured because they provide {\em semantic security}\footnote{The fact that no polynomial time probabilistic algorithm can derive information about a plaintext $m$ given its length, ciphertext and encryption algorithm, more than any other polynomial time probabilistic algorithm that has no access to the ciphertext.}, which is equivalent to {\em ciphertext  indistinguishability}\footnote{Given two plaintexts chosen by the adversary and the ciphertext of one of them chosen by us, the adversary cannot distinguish which of the plaintexts has been encrypted with a probability (significantly) larger than $1/2$, see \cite{FGsurveynonspc}.} \cite{GMsemanticEqualIndist,GsemanticEqualIndist}. 
This required randomness has a huge relevance in homomorphic encryption, as we will see. 

Encryption schemes are further distinguished by the relation between the encryption and decryption key. If the decryption key can be easily computed from the encryption one (in the typical case they are in fact identical), one speaks of a \textit{symmetric} cryptosystem, while if not, one speaks of an \textit{asymmetric} cryptosystem. Typical asymmetric systems also distinguish between  public (for encryption) and private (for decryption) keys $k_p, k_s$.

In modern cryptanalysis the adversaries are conceived as having finite computational resources and a cryptosystem is considered secure if its breaking is unfeasible with attack algorithms that are probabilistic in nature and running in polynomial time.
The running of the cryptosystems functions and adversary algorithms are all measured as a function of the so-called {\em security parameter} $\lambda$, which measures the complexity of the computational problem.


A cryptosystem is \emph{homomorphic} for an operation $*$ acting on $P$ if there is a corresponding operation $\circ$ acting on $C$ with
 \beq
   \texttt{Decr}(k_s, \texttt{Encr}(k_p, m_1) \circ \texttt{Encr}(k_p, m_2)) = m_1 * m_2 
 \eeq
for $m_1, m_2 \in P$.

Note this is not in general a true group homomorphism, as
 \beq
   \texttt{Encr}(k_p, m_1) \circ \texttt{Encr}(k_p, m_2)) \neq \texttt{Encr}(k_p, m_1 * m_2).
 \eeq
due to the (pseudo)randomness of the encryption scheme. However, while mathematically this lack of identity holds, there is a strong definition of homomorphic cryptosystems that reconciles with the group-homomorphism-like identity, in the statistical or computational senses, see \cite{HLHErv2017}. 

Defining a \emph{homomorphic encryption system} $(P, C, K, \texttt{Encr}, \texttt{Decr}, \texttt{Ev})$ then consists of specifying the evaluation function \texttt{Ev} that performs the homomorphically preserved operations $O$ on (a number of) ciphertexts
 \beq
   \texttt{Ev} : C^{n} \times \mathcal{C} \times K  \to C: (\vec{c}, O, p_k) \to c'
 \eeq
where $\mathcal{C}$ is the family  of circuits that the homomorphic cryptosystem can evaluate.
An homomorphic cryptosystem is defined {\em correct} if it correctly decrypts ciphertexts both coming from a circuit evaluation (sometimes called ``evaluated ciphertexts''), and from direct encryption of a plaintext (also dubbed ``fresh ciphertexts'').
Trivial homomorphic cryptosystems are excluded by requiring {\em strong homomorphicity} and {\em compactness} for the cryptosystems, for which we refer the reader to \cite{HLHErv2017}. We will consider exclusively non-trivial homomorphic cryptosystems.

Homomorphic cryptosystem can be further distinguished in 
\begin{itemize}
\itemsep=-1pt
 \item {\em partially homomorphic}: allow only one type of operation (addition or multiplication) for an \textit{unlimited} number of times,
 \item {\em somewhat homomorphic}: allow addition and multiplication, but only for a \textit{limited} number of times (the size of the ciphertext depends on the circuit depth),
 \item {\em levelled homomorphic}: allow addition and multiplication, but only for a \textit{limited} number of times specified as an input parameter (here the size of the ciphertext does not depend on the maximal allowed circuit depth, but the size of the public key does),
 \item {\em fully homomorphic} allow addition and multiplication for an \textit{unlimited} number of times, and thus arbitrary functions expressible as arithmetic circuits.
\end{itemize}

The randomness necessary for semantically secure schemes introduces a noise component that increases with each evaluation of an operation in the circuit. When the noise is above a certain limit, decryption is no longer correct.
Fully homomorphic systems, as constructed first by Gentry, see \cite{G2009}, can cope with this issue thanks to a procedure, called {\em bootstrapping} that allows to extend  specific somewhat homomorphic cryptosystems (called {\em bootstrappable}) to systems where unlimited number of operation evaluations are possible. Later realisations are, for example, 
\cite{SV2010, SS2010, CKLY2015,  BGV2012, FV2012, GSW2013, LTV2012}.
For the purpose of this work, it is important to note that the increase in noise is different for addition and multiplication. Typically the one induced by multiplications is much larger.

Moreover, and quite relevantly, in practical applications the computational complexity (and hence slowness) of homomorphic operations has  prompted to use \emph{levelled} homomorphic systems. This clearly affects the algorithms that can be successfully implemented at the practical level.

\section{\textbf{Approximation by Polynomials}}
By definition, somewhat, levelled and fully homomorphic encryption can only deal with polynomial operations, as they are the operations modelled by circuits of addition and multiplication. Therefore one takes the approach to approximate the functions one would like to evaluate homomorphically, with polynomials, e.g. see \cite{Hesa2018}.

For the convenience of the reader, we report here some fundamental elements of the theory of polynomial approximations, which will be relevant in the following.
It is well-known by the \emph{Weierstrass Approximation Theorem} (see e.g. \cite{Rudin}) that any real-valued continuous function $f$ on a closed interval $[a,b]\subset \mathbb{R}$ can be uniformly approximated by a polynomial, i.e. for every $\epsilon>0$ there exists a polynomial $p$ such that for all $x\in [a,b]$, $|f(x)-p(x)|<\epsilon$ or equivalently $||f-p||_{\infty}<\epsilon$, 
where $$||f||_{\infty}=sup\{|f(x)|: x\in [a,b]\}=max\{|f(x)|: x\in [a,b]\},$$ the \emph{supremum norm}. Moreover, there exists a unique polynomial of degree $n$ that minimizes the supremum norm within the set of all polynomials of degree $n$, which is called the
\emph{polynomial of best approximation} or \emph{minimax polynomial of degree $n$}. Being able to restrict the degree and to have control over the maximum error makes this type of approximation very attractive. However, as the supremum norm is not induced by an inner product, the theory of orthogonal projections cannot be applied, but luckily there exist several (numerical) algorithms, such as the classical \cite{Remez}, that can determine the minimax polynomial.

\section{Conditionals and Comparisons in Homomorphic Cryptosystems}\label{AnlResSect}
 
Homomorphic encryption aims at computing any computable function on encrypted data {\em without recurring to intermediate, not even partial, decryption}, and it has been highly regarded as a possibility to make privacy-safe machine learning algorithms. 
As mentioned in the introduction, homomorphic encryption 
allows to compute general polynomial operations, but in order to apply homomorphic encryption at least conceptually to 
general algorithms one still needs to prove, on the basis of the structured program theorem, that it can provide for comparisons as well as selections, the two fundamental components of conditionals.

We will take a completely novel approach and show that some of the involved aspects pose rather crucial problems for the program of homomorphic encryption.
Along the line of our proposed solution to implement comparisons and selections/jumps, we will also have to deal with other open issues in homomorphic encryption, such as the ability to perform divisions among ciphertexts.



\subsection{Implementing \textbf{Comparison Operations} in Homomorphic Cryptosystems}\label{ComparisonImplSect}


We start by defining a comparison operation as a map
\beq \label{CompFuncGeneral}
  \texttt{Comp}: C \times C \to \mathcal{S}=\{0, \pm 1\}
\eeq
where $C$ is the ciphertext space. We tackle the problem by trying to find a representation of this map in terms of elements of the circuit family that the homomorphic cryptosystem can evaluate, i.e. polynomial operations, rather than trying to implement comparisons via additional basic/elementary features of our cryptosystem (as attempted in OPE/ORE, and so far inconclusive). However, \texttt{Comp} is {\bf not} straightforwardly representable in terms of polynomials as it is discontinuous and typically implemented as a sign or equivalently using the  \emph{Heaviside (step) function}\footnote{We use the half-maximum convention.},
\[ H(x) =
\begin{cases}
1 & \text{if $x>1$} \\
\frac{1}{2} & \text{if $x=0$} \\
0 & \text{if $x<0$.}
\end{cases}
\]
Note that $H(x)=\frac{1}{2}(1+sgn(x))$.

Indeed, as $H$ is discontinuous, the Weierstrass Approximation Theorem 
does not apply and insisting on such an approximation requires using many polynomials of high degree, while the approximations are still of bad quality because of Gibb's phenomenon. As high polynomial order implies a high number of consecutive multiplications in the homomorphic system, this is problematic for the levelled or somewhat homomorphic schemes to which one is limited in practice as we have explained before.

We can however cope with these relevant issues and obtain a satisfactory definition and modelling.

\textbf{\textit{Solution to the problem}.}
We propose our solution, allowing: 1) to use only polynomial operations, 2) to compute comparisons in an efficient way in pure homomorphic encryption.

As we have remarked, \texttt{Comp} is typically implemented, as a sign or Heaviside function. Note that these are distributions, also called generalized functions\footnote{For an introduction see \cite{Kanwal}.}, and this allows us to base our solution on the representation of distributions as the {\em weak limit} of {\em sequences of locally integrable functions}.
This has the advantage that we can select suitable locally integrable functions admitting more convenient polynomial approximations that are amenable to homomorphic encryption.

Performing the weak limit is on the other hand problematic in the homomorphic encryption setting and in general when using (polynomial) approximations, which are typically defined only over restricted intervals.
We will solve this problem by selecting a class of locally integrable functions that have specific and suitable characteristics enabling us to calculate such limit in a sufficiently accurate way by mapping the values calculated over the restricted interval(s) to values at points outside such interval(s). A key-point will be keeping the number of consecutive operations sufficiently small (thus also keeping the necessary polynomials to be of a low degree).

After this general introduction to our solution, we now pass to its concrete illustration, in three key points hereby described in 1), 2) and 3).

\textbf{\textit{1) Choice of the sequence of locally integrable functions.}}

As we mentioned, it is well-known that $H$ can be obtained as the weak limit of several sequences of locally integrable functions, but in order to effectively perform the weak limit in homomorphic encryption with lower polynomials, we use the sequence
 \beq
  \{\tanh(kx)\}_k
 \eeq
where $\tanh(x)$ is the hyperbolic tangent, such that the weak limit becomes
  \beq \label{TanhHeavisideLimit}
  H(x) = \lim_{k\to\infty}{1 \ov 2}(1+\tanh(kx)).
 \eeq
The sequence of hyperbolic tangent functions will be crucial to allow us to effectively compute the weak limit in homomorphic encryption, as we will now explain.

Indeed, in homomorphic encryption we will have to polynomially approximate the functions $\tanh(kx)$. The approximation is necessarily only valid (that is, accurate) over a restricted interval. In fact for performance reasons in our case, the interval will be $[0, 0.25)$ in order to use the lowest possible order in polynomial approximations of the functions, as we will explain in point \emph{\textbf{3)}}. However, computing the weak limit means calculating the function over large intervals  defined by $z = kx, k \gg 1$.
We will manage to do so precisely because of the so-called \emph{bisection property} of $\tanh(z)$, which is why such sequence is crucial for us to be able map values calculated over the restricted interval to values at points outside such interval and obtain the weak limit rather efficiently (only low order polynomials will be necessary).

\textbf{\textit{2) Definition and calculation of the weak limit.}}

The weak limit in equation (\ref{TanhHeavisideLimit}) requires mapping the (approximate) calculated value of $\tanh(z)$ for $|z| \in [0, 0.25)$ to much larger $z$, effectively $z = kx, k \gg 1$ (as $k\to\infty$). In order to do so, as mentioned, we employ the \textit{bisection property}:
 \beq \label{BisectionTanh}
  \tanh(2z) = {2\tanh(z) \ov 1+\tanh^2(z)}.
 \eeq
After $r$ applications of this formula, the hyperbolic tangent initially calculated at $z=x$ is now calculated at $z=2^r x$, hence for $k = 2^r$ the limit $k \gg 1$ then corresponds to $r \gg 1$. 
We will discuss  later on what values of $r$ are achievable and/or efficient in practice, and what effect this has on the accuracy of the final comparison result.

 \textbf{\textit{The issue of divisions}.} Note that equation (\ref{BisectionTanh}) involves calculating a division, which is not possible in the present homomorphic encryption schemes. We solve this issue by using specific polynomial approximations for the function ${1 \ov x}$, where $x$ can then be generalised to functions of our ciphertexts, once we understand how to approximate the reciprocal function for a variable $x$. In the case of (\ref{BisectionTanh}), as for all $z$, $0\leq\tanh^2(z)\leq 1$, we must polynomially approximate the function ${1 \ov 1+x}$ for $x \in [0, 1]$ with $x = \tanh^2(z)$. We obtain the approximation of such function by shifting $x \to x+1$ in the approximation of the reciprocal function ${1 \ov x}$ for $x\in [1, 2]$. 

Obviously the higher the degree of the polynomial, the more accurate is the approximation, but, as said, we have to consider small degree polynomials because of the limitations of the levelled homomorphic cryptosystems we have to deal with in practice.
For illustration, let us consider the two lowest minimax
approximations (which can be found for example using the Matlab minimax algorithm):
\be
   {1 \ov x} & \approx 2.871320 - 3.029870x + 1.392785x^2 - 0.235498x^3\\
   {1 \ov x} & \approx 1.4571 - 0.5x  \label{DivisionLessOp}
\ee
both for $x \in [1, 2]$.
The first polynomial provides a better approximation (accuracy\footnote{The accuracy $\mu$ is related to the error $\epsilon$ as $\mu = -\log_2{\epsilon}$.} $\mu = 9.62$ bits, compared to 4.5 bits of the second one), but its degree is unfortunately bigger making it a less convenient candidate for homomorphic cryptosystems.

Coming back to equation (\ref{BisectionTanh}), we thus apply $x \to x+1$ in the approximation (\ref{DivisionLessOp}) of the reciprocal function ${1 \ov x}$ for $x\in [1, 2]$, which leads us to
\beq\label{DivisionLessOp01}
   {1 \ov 1+x} \approx 0.9571 - 0.5x  \qquad x \in [0, 1].
\eeq
Hence we can write the approximate bisection formula as
\beq\label{TanhBisec}
  \tanh(2z) \approx \tanh(z)(1.9142-\tanh(z)^2).
\eeq
 
\textbf{\textit{3) Polynomial approximation of the locally integrable functions.}}

We finally address the polynomial approximation of $\tanh(z)$ itself. Our choice for the explicit polynomial must also be guided by the fact that we are constrained in practice by the maximum number of consecutive operations that the levelled homomorphic system we are limited to can sustain before the need to bootstrap. Luckily, $\tanh(z) \sim z$ for $|z| \in [0, 0.25)$ with already quite good accuracy ($\geq 7.6$ bits). 

In practical applications with concrete datasets, this implies that datapoints must be preprocessed and in particular {\em normalised} such that the values we want to compare fall within the interval $[-0.12, 0.12]$ in order to apply the algorithms with the above described approximation. \textbf{To conclude:}
  
\begin{comp}
For $x_1, x_2\in [-0.12, 0.12]$, a comparison operator $$\texttt{Comp}(x_1,x_2)\in\{0,\pm 1\}$$ can be implemented under homomorphic encryption by approximating  the Heaviside function for $|x|<0.25$ with
$$\lim_{r\to\infty}{1 \ov 2}(1+\tanh(2^rx))$$
through iteratively replacing $\tanh(2x)$ by $x(1.9142-x^2).$ 
\end{comp}  

We detail the pseudocode of the comparison operation in  Algorithm \ref{CompAlgorithmEncr}.  

\subsection{Implementing \textbf{Selection/Jump Operations} in Homomorphic Cryptosystems}\label{SelectImplSect}

As we have mentioned above, the selection/jump operation is {\em particularly difficult in an homomorphic setting} and this is a crucial realisation of our analysis.
Indeed, public-key cryptography requires {\em ciphertext  indistinguishability}, which is evidently in tension with the necessity to select a path (one or more ciphertexts) at run time, that is, before the decryption, which is supposed to occur only at the end.

We propose, as best operational solution to this issue, an \textit{``implicit selection'' by weighting}. This is in fact not an actual selection so that it fully respects semantic security.
The idea is not to truly select, but to map the two subsets (the one of elements we ``want-to-select'', and the one of elements ``not-to-select'') into two different subspaces, choosing those spaces in a way that this map will keep them separate in the subsequent parts of any algorithm and will allow to recover at the end the ``want-to-select'' part. This is achieved by collapsing all elements of the ``not-to-select'' subset into the zero element of the ciphertext space, while the elements that we want to select will be preserved without change (that is, they will be mapped in themselves). We recall that in the case of homomorphic cryptosystems defined on polynomial rings the zero element is the zero polynomial.

The mapping procedure consists in re-scaling the compared data ciphertexts $x_1, x_2$ with suitable weights that depend on the result of the comparison.
There are different ways to implement such ``selection'' weights, differing in what comparison operation one wants to implement ($>$, $<$, \ldots) and what are the constraints on the number of consecutive operations. 


We now present implementations of comparison and a series of 
 {\em selection} operations, each of which can be realized as an algorithm in homomorphic encryption:
 \begin{select}
  \begin{align}
 & \texttt{Comp}: C \times C \to \{0, \pm 1\}, (x_1, x_2) \to  w_{12} \label{ComparisonAlgorithm} \\
 &  \texttt{Select}_{>_{1 \ov 2}}: C \times C \to C \times C, (x_1, x_2) \to  (s_{12}x_1, s_{21}x_2) \label{SelectMaggHalf} \\
 &  \texttt{Select}_{<_{1 \ov 2}}: C \times C \to C \times C, (x_1, x_2) \to  (s_{21}x_1, s_{12}x_2) \label{SelectMinrHalf} \\
 & \texttt{Select}_{=}: C \times C \to C \times C, (x_1, x_2) \to  (\overline{\overline{s}}x_1, \overline{\overline{s}}x_2) \label{SelectEQ} \\
 &  \texttt{Select}_{>}: C \times C \to C \times C, (x_1, x_2) \to  (\widetilde{s}_{12}x_1, \widetilde{s}_{21}x_2) \label{SelectMAGG} \\
 &  \texttt{Select}_{<}: C \times C \to C \times C, (x_1, x_2) \to (\widetilde{s}_{21}x_1, \widetilde{s}_{12}x_2) \label{SelectMINR}
 \end{align}
 with
$$ s_{ij} = {1+w_{ij} \ov 2}, \quad \overline{\overline{s}} = 1+w_{12}w_{21}, $$ 
 $$\widetilde{s}_{ij} = w_{ij}{1+w_{ij} \ov 2} \quad w_{ij} = -w_{ji}.$$
\end{select} 
Note that, although $w_{21} = -w_{12}$,  it is more convenient in the homomorphic cryptosystem scenario to calculate $w_{12}$ and $w_{21}$ independently, so that they have the same (lower) noise content, rather than $w_{21}$ having a higher one due to being the negation of $w_{12}$. This will improve accuracy and precision of the algorithms allowing more operations on the ciphertexts, but at the expense of time efficiency.

The $\texttt{Select}_{>/<_{1 \ov 2}}$ and the $\texttt{Select}_{>/<}$ algorithms differ in how they map the case $x_1 = x_2$: the former map $x_{1, 2} \to 0.5x_{1, 2}$, the latter map $x_{1, 2} \to 0$. Note that although the former algorithms do not implement exactly the $>$ and $<$ relations, they are convenient because they use less operations, and for some practical applications their treatment of the case  $x_1 = x_2$ is not very problematic.

Finally, in Algorithm  \ref{ComparisonAlgorithmEncr} we provide a detailed description using  $\texttt{Select}_{>_{1 \ov 2}}$ as an example (the algorithms for $\texttt{Select}_{>/<_{1 \ov 2}}$ and the $\texttt{Select}_{>/<}$ can be easily derived therefrom).

\begin{algorithm}
\caption{$\texttt{Comp}$, encrypted version.}\label{CompAlgorithmEncr}
\begin{algorithmic}
\Require Integer $r$ and encrypted $z_c = x_{c1}-x_{c2}$, where $x_{c1}, x_{c2}$ are encryptions of $x_{1},x_{2} \in [-0.12, 0.12]$ encoded using fractional encoder\\
\textbf{Constants} coefficient list $b_{\text{list}} = [-1.9142, 1.0, 0.5]$
\Ensure Binary values $\{0, 1\}$ with accuracy of about 3.65 bits
\State \textbf{Algorithm}
\For{$b \in$ coefficient list $b_{\text{list}}$} \textbf{do}
\State $b_e \gets \texttt{Enc}_f(b)$
\EndFor
\For{$i=0$ to $r$} \textbf{do}
\State Compute: $y_c \gets z_c*z_c$
\State Add plain: $u_c \gets -1.9142_e+y_c$
\State Multiply: $t_c \gets z_c*u_c$
\EndFor{}
\If{$r\%2 ==1$}
\State Negate: $z_c \gets -t_c$
\Else 
\State Assign: $z_c \gets t_c$
\EndIf \\
\Return $z_c = w_{12}$
\end{algorithmic}
\end{algorithm}


\begin{algorithm}
\caption{$\texttt{Select}_{>_{1 \ov 2}}$, encrypted version.}\label{ComparisonAlgorithmEncr}
\begin{algorithmic}
\Require Integer $r$ and $x_{c1}, x_{c2}$, which are encryptions of $x_{1},x_{2} \in [-0.12, 0.12]$ encoded using fractional encoder\\
\textbf{Constants} coefficient list $b_{\text{list}} = [-1.9142, 1.0, 0.5]$
\Ensure Binary values $\{0, 1\}$ with accuracy of about 3.65 bits
\State \textbf{Algorithm}
\State $s_{ij} = \texttt{Comp}(x_{ci}, x_{cj}; b_{\text{list}})$ for $ij = 12 \; \text{and} \; 21$
\State Add plain: $s_{ij} \gets s_{ij}+1.0_e$
\State \textbf{Multiply}: $s_{ij} \gets s_{ij}*0.5_e$
\Return $(s_{12}x_{c1}, s_{21}x_{c2})$
\end{algorithmic}
\end{algorithm}

We end this section with some comments on the specific features of the mechanism we have proposed to implement the selection operations.
First of all, we stress the main difference with an actual selection: while the latter operating on a certain set of elements returns in general a subset of it (typically, but of course not always, with fewer elements), our proposed mechanisms projects the unwanted elements onto the zero element, while preserving the elements one wants to select.

However, both the $w_{ij}$'s and the elements will be encrypted in the homomorphic case, thus we will not be able to discern which elements have been mapped to the zero element and which have been preserved (selected). Therefore, one will have to carry over all elements until the moment of decryption.
This clearly has implications for the efficiency of practical applications with large datasets. We explore some of the consequences of this in Section \ref{MachineLearningSect}.

We also observe that the final obtained values $s_{ij}, \overline{\overline{s}}_{ij}, \widetilde{s}_{ij}$ will never be exactly 1 or 0, but will tend asymptotically to those values.
An important figure of merit for the functions we have defined is the maximum number of consecutive operations they require, because the efficiency required in practical applications forces us to avoid bootstrapping, and thus allows only a limited number of consecutive operations.
We will discuss this in detail in Section \ref{ParmEmpiricEncrCompSect}.

\section{Tests and Results}\label{EmpiricalResultsComparisonSect}

\subsection{Methodology}\label{MethodsSect}

The general results presented in this work are \textit{agnostic} for what concerns the choice of (fully) homomorphic cryptosystem.
Nevertheless, to concretely implement our models and algorithms, we have chosen to adopt
the scheme of Fan and Vercauteren (FV) \cite{FV2012} for a series of reasons.
 \paragraph{\textbf{Efficiency}} the FV scheme is an efficient implementation of  the scheme in \cite{B2012}, one of the most remarkable second generation homomorphic systems.

 \paragraph{\textbf{Comprehension of the operating range for the cryptosystem parameters}}
determining the correct operating range of parameters for the various homomorphic cryptosystems is one of the active topics of research and it is unclear in all schemes. Other cryptosystems beside the Fan-Vercauteren one have been studied under this point of view, but their good parameter ranges are much less clear than the already incomplete one in Fan-Vercauteren's, as one can for example see mentioned and discussed in \cite{HS-EUROCRYPT2015}, see also \cite{CLP2017} when speaking of the popular scheme of \cite{BGV2012}. 
The FV scheme has been subject to a few more studies and experiments, as for example can be seen in the documentation of libraries such as {\tt SEAL} \cite{SEAL}, and \cite{BBBCIV2017}.

 \paragraph{\textbf{State of software libraries}} this is the point where the FV scheme is particularly valuable, with examples such as \cite{SEAL, A2014, FV-NFLlib}. In particular, {\tt SEAL} \cite{SEAL}
is evolving towards more explicit software engineering standards. We have been using its version 2.1, as latest updates have implemented modifications in the FV homomorphic scheme to improve the speed of calculations, but making it less simple to explore suitable ranges of parameters, see \cite{SEAL}. 

One important remark is that all libraries we know of do not actually implement the {\em fully} homomorphic cryptosystem, because they do not implement the bootstrapping, thus reducing the cryptosystem to only its somewhat homomorphic version. This will effectively limit the maximum number of consecutive operations we can evaluate. 

As our work is agnostic concerning homomorphic schemes, knowing the details of the FV one is not essential. It is however relevant to have a picture of the scheme's parameters, as they affect, for instance, the size of encodable data, the size of evaluable circuits, and so on. They are:
\begin{itemize}
  \item the plaintext modulus $t$, for the plaintext space $R_t \equiv {\mathbb{Z}_t[x] \ov f(x)}$
  \item the ciphertext modulus $q \gg t$, for the ciphertext space $R_q \equiv {\mathbb{Z}_q[x] \ov f(x)}$,
  \item the degree $d=2n$ of the monic irreducible polynomial modulus $f(x) = x^d+1$ (even degree and specific form chosen by {\tt SEAL} for efficiency reasons).
\end{itemize}
We will also encode data in plaintexts using the so-called \textbf{\emph{fractional encoding}} of \cite{DGLLNW2017}, by expanding our finite precision floats $u$ in a basis $b$ as $u = \sum_{i=0}^{u} u_i b^i + \sum_{j=1}^{s} u_j b^{-i}$, with $u_i, u_j$ then mapped to plaintext polynomial coefficients. This encoding depends on three parameters:
\begin{itemize}
 \item the basis $b$,
 \item the number of polynomial coefficients reserved for the fractional part $n_f = \text{max allowed}\,s$,
 \item the number of polynomial coefficients reserved for the integer part $n_i = \text{max allowed}\,u$.
\end{itemize}
 
\vspace{-.4cm}
\subsection{Datasets}\label{DtasetSect} 

As we have discussed in the introduction, only integers and finite precision floats (that is, rational numbers) are representable in the existing homomorphic cryptosystems. 
The datasets we have been using in our analysis are random datasets obtained from a uniform distribution of float values, and normalised according to the specifics we will illustrate in Sections \ref{ComparisonImplSect} and \ref{SelectImplSect}.

\vspace{-.3cm}
\subsection{Empirical Study}

We now turn to the empirical study of the algorithms leading to the functions in equations (\ref{ComparisonAlgorithm}
- \ref{SelectMINR}), which all depend on the single parameter $r$, the number of iterations to compute in the weak limit approximation derived from equation (\ref{BisectionTanh}).
We want  to determine for which values of $r$ and range of data arguments $x_1, x_2$ the algorithm is sufficiently accurate and this will involve studying the algorithms both in their unencrypted and encrypted form.


The tests are run over different datasets, specified in the respective subsections. The evaluation of the algorithm performances are based on the \textit{Mean Absolute Error} (MAE):
 \beq \label{MAEError}
  \text{MAE}(X, Y) = 
  \sum_{y \in Y} {|x-y| \ov |Y|} 
 \eeq
where $X$ is the set of expected values $x$, $Y$ the set of obtained ones (from the algorithms) and $|Y| = |X|$ denotes its cardinality.

\medskip

\subsubsection{\textbf{Evaluation of algorithm parameters - unencrypted form of the algorithm}\label{EmpiricUnencrCompSect}}

We first study the algorithms in unencrypted form to establish the dependence of the results on $r$.
We have considered a set $X_t$ of samples $x$ in the interval $|x|\in [0, 0.25)$ where the algorithms (\ref{ComparisonAlgorithm} - 
\ref{SelectMINR}) can operate. For each sample we have run the algorithm several times, for an increasing number of iterations $r$, starting from $r=1$. We have then evaluated the accuracy of the algorithm in the unencrypted form by calculating the MAE.

\begin{figure*}[ht!] 
\begin{center}
\includegraphics[height=0.4\textheight,width=\textwidth]{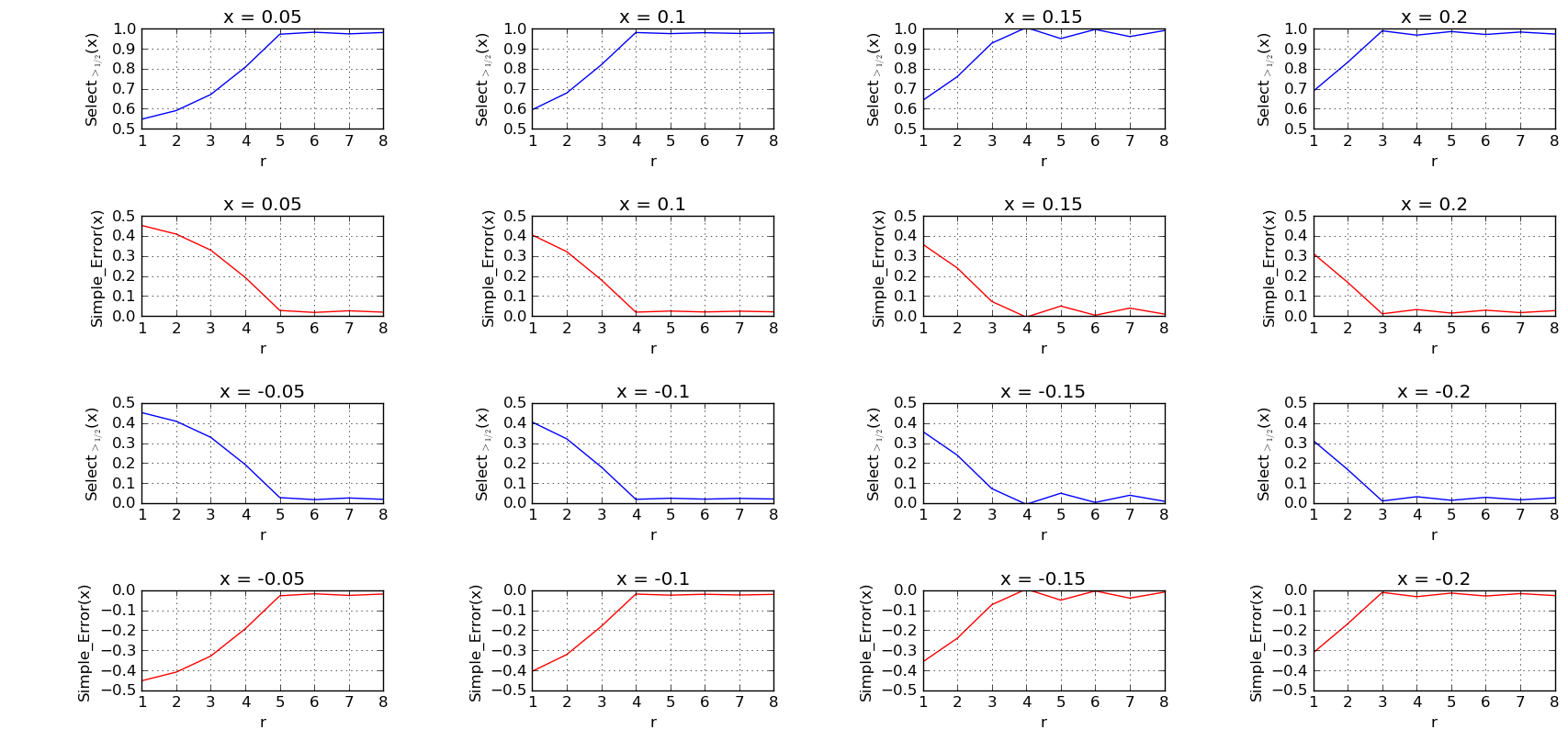}
\caption{Plots in rows one and three (blue color) show the value of the returned results from $\texttt{Select}_{>_{1 \ov 2}}(z, 0)$ against the number of iterations $r$. The plots in row two and four (red color) show the value of the simple error defined in Equation \ref{CompSimpleErro}.}
\label{ValuesAndErrorsComparisonAlgorithmUnencrypted} 
\end{center}
\end{figure*}

We present in Figure \ref{ValuesAndErrorsComparisonAlgorithmUnencrypted} a series of illustrative plots. We have chosen to report here plots concerning $\texttt{Select}_{>_{1 \ov 2}}(x, 0)$ with little loss of generality concerning the illustrative purpose, as those for $\texttt{Select}_{>}(x, 0)$ are quite similar, and as the algorithms for the $<$ operations are the same up to an intermediate sign. The plots in blue (rows one and three) show the returned results from $\texttt{Select}_{>_{1 \ov 2}}(x, 0)$ against the number of iterations $r$. The plots in red (row two and four) show the value of the simple error defined as
\beq\label{CompSimpleErro}
  \text{Simple\_Error}(x, r) = H(x)-\texttt{Select}_{>_{1 \ov 2}}(x, 0)
\eeq
where $H(x)$ is the Heaviside function. 
We have chosen to plot the results for some of the values $x$ we have considered. In particular we plot for points with values $x$ going from $-0.20$ to $0.20$ in steps of $0.05$ in order to provide a consistent coverage of the working interval of the algorithms.

We also present in Table \ref{TableErrorComp} the values of the MAE, equation \ref{MAEError}, for those set of values for $x$ and for the tested number of iterations $r$ in the cases of $\texttt{Select}_{>_{1 \ov 2}}(x, 0)$ and $\texttt{Select}_{>}(x, 0)$.
The results in the table show that, in the former, for $r=4$ the MAE has dropped at around $6\%$, and for $r=5$ around $3\%$, while in the latter the values are slightly bigger. The low degree of the approximating polynomial from equation (\ref{TanhBisec}), 
and thus low number of necessary consecutive multiplications, make this an interesting result for applications in homomorphic encryption.

\begin{table*}[t]
\center{
\begin{tabular}{|c|c|c|c|c|c|c|c|c|}\hline
  & $r=1$ & $r=2$ & $r=3$ & $r=4$ & $r=5$ & $r=6$ & $r=7$ & $r=8$\\ \hline\hline
$\text{MAE}(r) \;\text{for}\;$ $\texttt{Select}_{>_{1 \ov 2}}$ & $0.38$ & $0.28$ & $0.15$ & $0.062$ & $0.027$ & $0.017$ & $0.026$ & $0.019$ \\
\hline
$\text{MAE}(r) \;\text{for}\;$ $\texttt{Select}_{>}$ & $0.47$ & $0.39$ & $0.22$ & $0.10$ & $0.056$ & $0.034$ & $0.051$ & $0.036$ \\
\hline
\end{tabular}
}
\vskip 0.25cm
\caption{Values for the mean absolute error defined in Equation \ref{MAEError} for a number of iterations values $r$ calculated over a set of points $ x_t = 0.05*s$, $-4 \leq s \leq 4$.}
\label{TableErrorComp}
\end{table*}

\medskip

\subsubsection{\textbf{Selection of algorithm and homomorphic scheme parameters - encrypted form of the algorithm}}\label{ParmEmpiricEncrCompSect}

We move now to a series of tests with a carefully chosen artificial dataset to establish in the best $r$ and FV cryptosystem parameters, where {\em best} means leading to the smallest errors
over the maximum possible data-value interval.

The value of $r$ determines the number of consecutive operations the algorithms {\em must} sustain, while the parameters of the homomorphic scheme determine the number of consecutive operations the {\em encrypted} algorithm {\em can} sustain without bootstrapping.
In the case of algorithms (\ref{SelectMaggHalf}
- \ref{SelectMINR}), the number of (noise-dominating) consecutive multiplications  we need to be able to perform to run $r$ iterations is
 \be
       & 2 r + 1_p && \quad \text{for the} \quad \texttt{Select}_{>/<_{1 \ov 2}} \quad \text{algorithms} \\
       & 2 r + 1+ 1_p && \quad \text{for the} \quad \texttt{Select}_{>/</=} \quad \text{algorithms}
 \ee
where $1_p$ is a multiplication with a plaintext coefficient\footnote{We distinguish mixed plaintext/ciphertext multiplications because the noise level estimates are different than pure ciphertext multiplications in our implementation based on {\tt SEAL}, see Table 3 in \cite{CLP2017}.}, and $2r$ or $2r+1$ are ciphertext multiplications. The total count of operations is  of course higher, when including additions and relinearizations, but as they generate less noise, we will neglect them. Moreover, note that the ciphertext multiplications involve multiplying recursively the {\em same} ciphertext, which means that 
successive multiplications are more costly for the noise growth, as they involved already noise-grown ciphertexts.

A number of consecutive operation such as, for instance,
8+1 (for $r=4$) may seem not huge, but it must be put in relation with the cryptosystem parameters necessary to accommodate it.
As said before, the analysis concerning the choice of parameters is still an active field of research, and there are in fact different partial results in the literature.
For example \cite{CLWW2015} estimates the cryptosystem parameters for a scheme like the FV one, finding that already to perform 10 multiplications (of different, and thus with minimal noise, ciphertexts) requires a polynomial modulus degree of $d = 8192$ and a plaintext modulus of at least $2^{243}$ for a fractional encoding in base $b=3$, which in turns implies a value for the ciphertext modulus of about $2^{226}$ to have 123 bits security as estimated in \cite{CLP2017}.
However, the work \cite{AEH2015} claims that much higher values are actually necessary to be able to perform 4 subsequent multiplications, already in the case of a much simpler integer encoding ($t = 131702, \; q > 2^{159}, \; d = 8192$0. Finally, the recent paper \cite{BCIV2016} claimed necessary values of the form $t \gtrsim 2^{107}, d \leq 368$ when dealing with the case of a graph (representing an instance of Ivakhnenko's group method of data handling), whose evaluation along a path from input to output comported $\approx 6$ consecutive multiplications (plus a similar quantity of additions).

In summary, two things appear from the literature:
\begin{itemize}
\item the parameter choice bounds are coarsely estimated for similar experiments,
\item the number of consecutive multiplications implemented in existing literature is very low, thus our $\sim 8+1$ one appears to be the highest ever tried.
\end{itemize}

We now present the results for the algorithms, $\texttt{Select}_{>/<_{1 \ov 2}}$ and $\texttt{Select}_{>/<}$ (given the number of operations required by these, the results also apply to the case $\texttt{Select}_{=}$).
We have been testing with parameters ranging over a number of possible values, in particular
 \beq \label{ParametersRange}
 \begin{array}{ll}
  d & \in \{8192, 16384, 32768\} \\
  q & \in \{2^{116} - 2^{18} + 1,  2^{226} - 2^{26} + 1,  \\
   &   \qquad{} 2^{435}- 2^{33} + 1, 2^{829} -2^{54} - 2^{53} - 2^{52} + 1\} \\
  t & \in \{4096, 16384, 65536\} \\
  b & \in \{3, 5, 7, 9\} \\
  n_f & \in \{6, 8, 10, 24, 32\} \\
  n_i & \in  \{8, 16, 32, 64\}
 \end{array}
 \eeq
where the parameters and their notation have been defined in Section \ref{MethodsSect}.
The time of key generation and storage overhead following from these choices of parameters are known for the FV scheme and the {\tt SEAL} implementation library, which we have used in our tests, and we refer the reader to the literature, see the original articles \cite{FV2012, SEAL}. We also recall, and stress, that our algorithms are agnostic for what concerns the choice of homomorphic scheme, and thus any performing scheme may be used in practical applications, so that computation and storage overheads can be tuned as desired.

The range of values for $q, t, d$ were chosen by taking advantage of the sets of values indicated by the {\tt SEAL} team in their testings of the library versions 2.1 and 2.2. Instead, {\tt SEAL} version 2.3.0 uses a modification of the FV scheme to increase time efficiency, but which allow somewhat less flexibility and ``ease" in the choice of parameters. In particular, the available values of the parameter $q$ in version 2.3.0 have proven in our case to yield sub-optimal results.

We have run the algorithm $\texttt{Comp}(z, 0)$ over a small dataset 
$$ z \in \{ -0.20, -0.15, -0.10, -0.05, 0, 0.05, 0.10, 0.15, 0.20\}$$
capable however to cover sufficiently uniformly the allowed instance space $|z| \in [0, 0.25)$ from Sections \ref{ComparisonImplSect} and \ref{SelectImplSect}.

We list in Table \ref{TableResultsCompEncr} the best results for each tested value of $r$, where best indicates smallest MAE for the selection weights $s_{ij}, \overline{\overline{s}}_{ij}, \widetilde{s}_{ij}$ as defined in Equation (18).

\begin{table*}[t]
\center{
\begin{tabular}{|c|c|}\hline
  Iterations & Results \\ \hline\hline

\multirow{2}{*}{$r=3$} & Smallest parameters where result still achieved: \\
      & $ d=16384, \; q = 2^{435} - 2^{33} + 1, \; t = 65536, \; b =7, \; n_i = 8, \; n_f = 8 $ \\
      \hline
\multirow{4}{*}{$r=4$} & Not accomplished correctly (error less than 1 at least) \\
      & by any parameter value in \ref{ParametersRange}. ``Best results'' for \\
      & $ d=16384, \; q = 2^{435} - 2^{33} + 1, \; t = 65536, \; b =7, \; n_i = 8, \; n_f = 8 $ \\
\hline
\end{tabular}
}
\vskip 0.25cm
\caption{Values for the mean absolute error defined in Equation \ref{MAEError} for a number of iterations values $r$ calculated over a set of points $ x_t = 0.05*s$, $-4 \leq s \leq 4$.}
\label{TableResultsCompEncr}
\end{table*}

From our analysis, the rationale behind the effectiveness of encryption scheme parameters emerges as follows. First of all we need small $t$ and large $q$ since ${q \ov t}$ mostly determines the maximum noise bound, see \cite{CLP2017}. Secondly, we need to keep the number of coefficients reserved to the fractional part in encoding ($n_f$) as small as possible because  during multiplication the number of coefficients occupied by the fractional part will increase rapidly. The number of coefficients reserved to the integer part ($n_i$) is of less concern, because all the normalised test data instances $x$ are smaller than one. 

The basis $b$ used for the fractional encoding, see Section \ref{MethodsSect} also played an important role. One would like to have as small a basis as possible, to avoid the ``wrapping up'' of the modulo $t$ during computations. However, smaller basis also means that more coefficients of the plaintext polynomial will be non-zero, and so since the number of coefficients of the fractional part increase with multiplications, they can more easily cross over to the coefficients reserved for the integer part, and ruin decryption. 

Finally, the degree $d$ of the polynomial modulus is relevant because of two different reasons: on the one hand the experiment should take into account security bounds, which depend on $d$ since long polynomial are more difficult to attack; on the second hand having a big number of coefficients also helps in avoiding that those reserved to the fractional part and those to the integer part cross over and mix up rapidly.

\subsubsection{\textbf{Full tests - encrypted form of the algorithm}}\label{FullEmpiricEncrCompSect}

Having estimated as discussed the best algorithm and scheme parameters, we have finally run full-fledged tests over randomised datasets to assess the accuracy of the algorithms.

We have studied the algorithms $\texttt{Select}_{>_{1 \ov 2}}$,  $\texttt{Select}_{>}$ and $\texttt{Select}_{=}$, since 
the algorithms for the $<$ (less than) relations are essentially the 
same as the ones for $>$ up to an (intermediate) sign, hence the test results apply to those as well.
Our datasets consisted of couples of datapoints randomly generated in the range $[-0.12, 0.12]$ (so that the difference between datapoint values would fall in the valid range $[0, 0.25)$ to apply the algorithms, see Sections
\ref{ComparisonImplSect} and \ref{SelectImplSect}).
We have used several measures of accuracy and performance for the algorithms, to be able to provide a rigorous evaluation.

The results are presented in Table \ref{TableFullAccuracyResultsAlgorithms}. The simplified notation
MAE($a$) indicates the error calculated using equation (\ref{MAEError}) on the values $a = a_{\text{out}}-a_{\text{expected}}$. We have studied various tests, in particular we have considered $a$ to be first $s_{ij}, \overline{\overline{s}}_{ij}, \widetilde{s}_{ij}$ and then the final full output of the algorithms (that is, $a = s_{ij}x_i$ and the analogous for $\overline{\overline{s}}_{ij}, \widetilde{s}_{ij}$).

The best performing algorithms are $\texttt{Select}_{>_{1 \ov 2}}$ (and thus $\texttt{Select}_{<_{1 \ov 2}}$ as it is the same up to an initial sign), achieving about 20\% error on the selection weights and 2\% on the final 
output. The error is dominated by the error value for datapoints that are very close to each other. In fact, we have run the same tests with datapoints with a fixed minimal distance in order to check variations depending on this, and the error rate drops rapidly in function of the inter-distance of points (already with inter-distance higher than a few percent, for instance 3\%, the error rate on selection weights drop at about 12\%).

The algorithm $\texttt{Select}_{=}$ deserves a special comment: the comparison weights $w_{ij}$ are exactly 1 when $x_i = x_j$ and different from 1 when $x_i \neq x_j$ (and closer to 0 as the difference/distance between $x_i$ and $x_j$ is larger), so that if in a simple application one lets the data owner simply decrypt the comparison weights and pick the datapoints corresponding to weights equal to 1 to perform the selection part of the conditionals of interest, 
one would have perfect accuracy.
This however cannot be done when the algorithm must be inserted in a longer pipeline of algorithms and the ``selection'' must be performed on the encrypted parts and carried over to further steps of the pipeline. The results relative to $\texttt{Select}_{=}$ that we show in Table \ref{TableFullAccuracyResultsAlgorithms} are therefore to be intended for this case.

\begin{table*}[t]
\center{
\begin{tabular}{|c||c|c||c|c||c|c||}\hline
\multicolumn{7}{|c|}{$d=16384, \; q = 2^{435} - 2^{33} + 1, \; t = 65536, \; b =7, \; n_i = 8, \; n_f = 8 $}  \\ \hline\hline
  Iterations & \multicolumn{2}{|c||}{$\texttt{Select}_{>_{1 \ov 2}}$} &  \multicolumn{2}{|c||}{$\texttt{Select}_{>}$} &  \multicolumn{2}{|c||}{$\texttt{Select}_{=}$} \\ 
\hline
     & MAE($s_{ij}$) & MAE($s_{ij}x$) & MAE($\overline{\overline{s}}_{ij}$) & MAE($\overline{\overline{s}}_{ij}x$) & MAE($\widetilde{s}_{ij}$) & MAE($\widetilde{s}_{ij}x$) \\
\hline
$r=3$ & 0.26 & 0.021 & 0.41 & 0.023 & 0.52 & 0.057 \\
\hline
$r=4$ & 1.7 & 0.28 & 4.9 & 0.55 & 2.6 & 0.30 \\
\hline
\end{tabular}
}
\vskip 0.25cm
\caption{Errors for the comparison and selection/jump 
algorithms defined in (\ref{ComparisonAlgorithm}), (\ref{SelectMaggHalf}), (\ref{SelectEQ}), (\ref{SelectMAGG}). We have tested the algorithms on randomly generated datasets with batches of 60 couples of datapoints to be compared and present here the average results for $r=3$, while for $r=4$ we present the result for the best batch (since anyway the case $r=4$ is affected by error of decryption due to too many consecutive operations performed, see Section \ref{EmpiricalResultsComparisonSect}.}
\label{TableFullAccuracyResultsAlgorithms} 
\end{table*}


\begin{table*}[t]
\center{
\begin{tabular}{|c|c|c|c|}\hline
\multicolumn{4}{|c|}{Average timing per instance in seconds} \\
\hline
\hline
\multicolumn{4}{|c|}{$d=16384, \; q = 2^{435} - 2^{33} + 1, \; t = 65536, \; b =7, \; n_i = 8, \; n_f = 8 $}  \\ 
\hline\hline
  Iterations & $\texttt{Select}_{>_{1 \ov 2}}$ &  $\texttt{Select}_{>}$ &  $\texttt{Select}_{=}$ \\ 
\hline
$r=3$ & 17.4 s & 21.5 s & 21.1 s \\
\hline
$r=4$ & 30.5 s & 31.5 s & 31.2 s \\
\hline
\end{tabular}
}
\vskip 0.25cm
\caption{Values for the timing for runs of the selection algorithms in seconds per instance. }
\label{TableTimingResultsCompEncr}
\end{table*}

We finally present in Table \ref{TableTimingResultsCompEncr} the result for the timing of the selection/jump algorithm. 
Our work has not focused on achieving the best performance, as it has been more centred on the proof-of-concept and the practical implementation of the algorithms, as well as to the discussion of the novel issues concerning homomorphic encryption and applications such as machine learning (see Section \ref{MachineLearningSect}).
We have however measured the timings when running our tests, and report in the table the average timing per $(x_1, x_2)$ instance for the algorithms (\ref{SelectMaggHalf}), (\ref{SelectEQ}), (\ref{SelectMAGG}).
A direct comparison with the results reported in the literature is however not straightforward, because:
\begin{itemize}
 \item there are very few works implementing similarly complex algorithms in a homomorphic cryptosystem,
 \item often the experiments in the literature have been performed on powerful computer clusters, see e.g. \cite{AEH2015},
 \item only few among the works with complex algorithms report full algorithm timings\footnote{The others,  e.g. \cite{BCIV2016}, report ``time per operation'' such as addition, multiplication or encryption. However, also other routines such as relinearizations are present and finding the overall algorithm timing is not straightforward.}.
\end{itemize}

\vspace{-.6cm}
\subsection{Improvements and comments }

We have presented here above a series of algorithms to evaluate comparisons and conditionals in homomorphic encryption settings.
The algorithms have been explicitly tested in a concrete implementation of the Fan-Vercauteren encryption scheme in a levelled form.
The limitation on the total number of consecutive operations has the strongest influence on the accuracy of the algorithms. 

Such limitations, and hence inaccuracies, would simply be absent for an implementation in a fully homomorphic scheme, or, possibly,
using schemes that although limited can tolerate a larger number of consecutive operations (we estimated 9 multiplications would already guarantee accuracy at percent or sub percent level).

It would be also interesting to assess what effects new plain- and ciphertext encodings such as \cite{BBBCIV2017} would have, possibly in alleviating some of the accuracy loss due to crossing of the integer and fractional parts of the standard encoder, see Section \ref{FullEmpiricEncrCompSect}.

\section{Applications}\label{ApplicationsSect}

\subsection{Machine learning and specific issues}\label{MachineLearningSect}

As mentioned in the introduction, part of the present interest in homomorphic encryption stems from the potential to permit privacy preservation to coexist with the nowadays ubiquitous machine learning/data mining/predictive analytics\footnote{Again, for brevity of expression in the following we will use ``machine learning'' as an umbrella term for all these different but related approaches.} without incurring in the limits of the privacy/data usefulness budget of other approaches \cite{FLJLPR2014}.

In the literature of the past few years we can find a limited number of implementations of machine learning algorithms\footnote{Typically for prediction only, that is having the training part all in unencrypted form.} preserving privacy combined with homomorphic cryptography techniques, see e.g. \cite{GLN2012, BBBCIV2017,  BLN2014, GDLLNW2016, AEH2015}. 

The main problem is that very often machine learning techniques are not amenable to homomorphic encryption due to various limitations and it is therefore interesting to re-think the actual machine learning algorithms.

\subsubsection{\textbf{Relevant aspects of machine learning (ML)}}
\mbox{}


Developing machine learning/data mining/predictive analysis systems typically involves a series of different steps\footnote{Not always: for example instance-based methods, such as $k$-nearest neighbours, do not need an actual training, validation and testing phase.}:
training,
validation,
testing,
prediction.
The core issue is coined as solving a ``learning problem'', which comes down to finding within a certain solution space (essentially delimited by the inference bias) a function or generalisation thereof that maps input data to a correctly inferred output (be it classification or regression). To this end, the algorithm uses the input data to assess the relevance of different hypothesis in the solution space, building them against the available training data, and validating and testing them against independent pieces of data. The tested algorithm can then be used with other data for prediction.

\subsubsection{\textbf{Problematic points of ML in homomorphic settings}}

We will now elaborate on certain ingredients of the machine learning inference process that clash in a particularly relevant way with fundamental constraints of homomorphic encryption. 
In a large class of machine learning algorithms two element are paramount: the {\em stopping criterion} and {\em heuristics}. 

\begin{itemize}

\item \textbf{Stopping criterion}. Typically machine learning algorithms terminate when a stopping criterion is met, generally when an extremal condition is reached. 
This means that the algorithm must be able to evaluate a condition (the criterion) and select one of the options (essentially, continue or stop) while running in its encrypted form.


In Section \ref{SelectImplSect} we explained that the selection step conflicts with the fundamental requirement of semantic security in (homomorphic) encryption and we proposed a  ``selection by weighting" algorithm that allows mapping the selected-for and unselected-for subsets to specific subsets (in particular the zero
element for the unselected subset) that at decryption will provide the desired result.
  
  The bigger problem in this case is that the ``selection by weighting" does not really signal that a selection has been made, but all mapped results (both ``selected'' and ``rejected'' ones) are still encrypted and carried over.
There is no way at run time to determine that the stopping criterion has thus been met and the procedure must be stopped, until decryption occurs.
Unfortunately, and importantly, not stopping the training of an algorithm precisely at the stopping point does
  entail overfitting and thus suboptimal learning models.

\item \textbf{Heuristics}.  In order to efficiently explore the instance (data) and problem space, and make useful inference, several machine learning algorithms operate heuristic choices at run time. Again, the clash between the need to make selections and the requirement of semantic security of the (homomorphic) encryption pops up. Differently from the case of the stopping criterion, here our ``selection by weighting'' would not create loss of accuracy in the algorithms, but, of course, in the case of large datasets it would entail carrying over the full dataset all along, hence affecting the efficiency of the algorithm, and in certain cases its whole inference capabilities, as we will discuss at the end of this section. 

Also another issue can arise: the comparison weights $w_{ij}$ are not exactly 1 or 0 but some other numbers (float) close to that, because of the limitations in the number of consecutive operations of the levelled homomorphic system one has to use in practice, which limits the value of the algorithm parameter $r$ and thus its accuracy. This effectively transforms a machine learning algorithm into a weighted version of itself. In some cases this does not significantly affect the accuracy, sensitivity and precision of the algorithms, but in other cases it does, also in an adverse way. The studies in this respect are scarce in the literature, see e.g. \cite{ABBL2012} for what concerns clustering.
\end{itemize}
The two issues here presented are quite fundamental and could seriously complicate, if not make impossible, the implementation of privacy-preserving machine learning using purely homomorphic encryption. The stopping criterion issue, in particular, implies that the training of correctly performing algorithms (that is, not overfitted ones) does not seem achievable without decryption at run time, which goes against the  aim itself of homomorphic encryption. While this drawback affects only the training of models, private data are also used in training (even more than in the prediction runs after training) and should therefore be protected as well.

The heuristics issue instead seems to represent a secondary problem, only affecting performance and thus possibly solved by, for example, hardware evolution. However, that is not the case. To be able to make actual inferences several machine learning algorithms do need to operate with heuristics on the data/problem space. If that is not possible, the algorithms cannot proceed with meaningful inferences. Again, this appears to be a relevant obstacle on the road to make machine learning privacy-friendly by using homomorphic encryption.

\subsection{Applications to algorithms different than machine learning}\label{ApplDiffSect}

Algorithms different from machine learning or similar predictive analytic techniques that do not need to make inferences and avoid overfitting as discussed in the previous section, are not in such a relevant clash on general grounds with homomorphic encryption. 

This means that operations such as pure database searches, for instance, even including comparisons, conditionals and selections could be effectively performed taking advantage of the techniques and algorithms we have developed in this work and their future improvements.

\section{Conclusions}\label{ConclusionsSect}

Homomorphic encryption provides in theory an elegant solution to the problem of privacy preservation in data-based applications, such as those provided and/or facilitated by machine learning techniques, but several limitations hamper its implementation. 
In this work we have identified, on the basis of the structured program theorem, the set of minimal operations that guarantee the computation of any computable function or algorithm.
We have then focused on those that are still lacking in homomorphic encryption, namely comparisons and conditional selections.
We have discovered rather fundamental clashes between the necessity to implement those operations and the basic requirements of (homomorphic) encryption.
We have also proposed practical implementations for those operations or their closest possible forms in homomorphic encryption. The limitation on the total number of consecutive operations, due to the use of levelled homomorphic encryption schemes without using bootstrapping (a practical limitation we have to face), has had the strongest influence on the accuracy of our algorithms. Percent accuracy (and better) can be obtained however for datapoints which are sufficiently inter-spaced.
Moreover, such limitations, and hence inaccuracy, would not occur in a fully homomorphic scheme, or, possibly,
using schemes and/or encodings that can tolerate a larger number of consecutive operations even if only somewhat homomorphic (we estimate from 9 multiplications onward).

We have also analysed the specific situation arising in machine learning/predictive analytic applications. 
We have pointed out at least two main sources of tension with the use of homomorphic encryption to fully guarantee privacy preservation in machine learning due to the newly found above-mentioned issues.
These two sources of tensions are the stopping criterion and heuristics. They are present and paramount in most machine learning algorithms, and clash with (homomorphic) encryption in that they require performing selection/jump operations at run time, which in its turn clashes with semantic security, as we have studied in this work.

Two options are open under this respect. On the one hand it might be possible to find new classes and families of learning algorithms that operate without ``choices at run time''. On the other, we could reconsider the use of homomorphic encryption. Maybe some other technology, such as for example functional encryption (e.g. \cite{GGHRSW2013}) may be capable to avoid the need for high level of communication and intermediate decryption of other techniques (such as secure multi-party computation ones)? Functional encryption could allow to limit the computations to some agreed level, while preserving the rest of the privacy of the data or algorithm. 

Note however that the class of algorithms which do not need to make inferences and where overfitting can be avoided, are not in conflict with the general grounds of homomorphic encryption. 
We believe that further exploring the use of homomorphic encryption in algorithms for privacy preservation is paramount.



\end{document}